\documentclass{moriond}

\usepackage{comment}
\usepackage{braket}
 \usepackage{amsmath}
 \usepackage{amssymb}
\def\Journal#1#2#3#4{{#1} {\bf #2}, #3 (#4)}

\def\PRD{{\em Phys. Rev.} D}

\def\be{\begin{equation}}
\def\ee{\end{equation}}
\def\bea{\begin{eqnarray}}
\def\eea{\end{eqnarray}}

\renewcommand{\d}{\mathrm{d}}

\newcommand{\Photo}{}

\begin{document}
\vspace*{4cm}
\title{Axion-like Dark Matter Mediators}

\author{Sophie Mutzel}


\address{Aix Marseille Univ, Universit\'{e} de Toulon, CNRS, CPT, iPhU, Marseille, France}

\maketitle
\abstracts{
During the last decades, experimental advances have significantly constrained the standard electroweak-scale WIMP produced via thermal freeze-out, leading to a shift away from this standard paradigm. Here we explore the possibility of an axion-like particle (ALP), the pseudo-Goldstone boson of an approximate U(1) global symmetry spontaneously broken at a high scale $f_a$, acting as a mediator between the Standard Model (SM) particles and the dark matter (DM) particles. We focus on the case where the couplings are too small to allow for DM generation via freeze-out and the DM is thermally decoupled from the SM particles. However, alternative mechanisms like freeze-in and freeze-out from a decoupled dark sector can still reproduce the observed DM relic density. Having determined the region of parameter space for these scenarios, we then revisit experimental constraints on ALPs from electron beam dump experiments, astrophysics and rare B and K decays.
}
\section{Introduction}
Recent experimental advances now severely constrain models in which the DM particle is directly charged under the SM gauge group via renormalisable interactions. The standard WIMP produced via thermal freeze-out is, for instance, challenged by the fact that - despite experimental searches for decades - it has not been detected so far. Hence, one might wonder if different scenarios and different candidates are also possible. In fact, the QCD axion is itself a pioneering example of a non-thermally produced DM candidate. In this work, the axion does not constitute the DM but we explore the possibility of a DFSZ-like ALP acting as a mediator between the DM particle and the SM. The size of the new couplings is driven by the new symmetry breaking scale $f_a$ which should be large enough to be in accord with collider experiments today. Therefore the new couplings are small and we focus on regions where the DM particles are out-of-equilibrium from the SM particles.
\section{The Model}
In our model, we extend the SM by a Dirac fermion $\chi$ and a DFSZ-like ALP $a$, by which we mean the pseudo-Goldstone boson of an additional approximate $U(1)$ global symmetry spontaneously broken at a high scale $f_a$. The ALP acts as a mediator between the SM and the dark sector. Below the electroweak scale, the ALP couples effectively only to SM fermions at tree-level via Yukawa-like interactions and interacts with gauge/Higgs bosons via fermion loops. If there are no heavy vector-fermions integrated out, i.e. if the fermion content comes only from the SM, the anomaly cancellation requirements automatically imply that the contributions from dimension-5 terms cancel to leave only dimension-4, mass-dependent terms \cite{axionanomalies}. We hence consider the following relevant new terms in the effective Lagrangian \footnote{Note that in some UV complete model where the ALP emerges from an extended Higgs sector one also expects dimension-5 couplings between the ALP, the SM fermions and the Higgs which give ultraviolet sensitive contributions to the dark matter relic density.}
\begin{eqnarray}
\mathcal{L} \supset\frac{1}{2} \partial_\mu a \partial^\mu a - \frac{1}{2}m^{2}_{a} a^2+ \sum_f \frac{m_f}{f_a}C_f \bar f i \gamma_5 f  + \frac{m_\chi}{f_a} C_\chi \bar \chi i \gamma_5 \chi a \; . \label{eq:lagrangian}
\end{eqnarray}
Here, $f$ is any Standard Model fermion with mass $m_f$, the masses of the $\chi$ and of the ALP are $m_\chi$ and $m_a$, respectively. Furthermore, we define $g_{a\chi\chi} \equiv m_\chi  C_\chi/f_a$ as the \emph{hidden sector coupling} and $g_{aff} \equiv C_f/f_a$ as the \emph{connector coupling}. This Lagrangian is valid below the electroweak symmetry breaking scale. 
\section{Dark Matter Genesis Scenarios}
Deducing the interactions amongst the SM, DM particles and the ALPs from the Lagrangian in eq.~\ref{eq:lagrangian} allows us to write down the most general, complete set of coupled differential equations governing the evolution of the $a$ and $\chi$ number densities
\begin{equation}
\renewcommand{\arraystretch}{1.3}
\begin{array}{rc@{\,}c@{\,}l}
\frac{\d n_{\chi}}{\d t}+3 H n_{\chi}&&=&\sum_f \left\langle\sigma_{\chi\bar \chi\rightarrow f \bar f}v\right\rangle\left( n_{\chi}^{\rm{eq}}(T)^2-n_{\chi}^{2}\right)+\left\langle\sigma_{a a \rightarrow \chi\bar\chi } v\right\rangle n_{a}^{2}-\braket{\sigma_{\chi\bar\chi \rightarrow a a} v}n_{\chi}^2\,,\\
\frac{\d n_a}{\d t}+3 Hn_a &&=& \sum_{i,j,k}\left\langle\sigma_{i a \rightarrow j k} v\right\rangle\left(n_{a}^{\rm{eq}}(T) n_{i}^{\rm{eq}}(T)-n_{a} n_{i}^{\rm{eq}}(T)\right)+ \braket{\Gamma_{a}}\left(n_{a}^{\rm{eq}}(T)-n_{a}\right)\\
\label{eq:generalBoltzmann}\,\, &&- & \braket{\sigma_{a a \rightarrow \chi\bar\chi} v}n_{a}^2+\braket{\sigma_{\chi\bar\chi \rightarrow a a} v}n_{\chi}^2 ,
\end{array}
\end{equation}
with $i,j,k$ SM particles which participate in the ALP number changing processes (the dominant contribution coming from the t-channel process $i=g$, $j=\bar k = t$). 
The $2\to 2$ processes enter via the typical thermally averaged cross section  $\braket{\sigma v}$ and the (inverse) decay of the pseudoscalar via the thermally averaged decay rate, $\braket{\Gamma_a}$ \cite{cosmicabundances}. We have applied the principle of detailed balance where appropriate.
\begin{figure}[t]
\begin{center}
	\includegraphics[scale=0.6]{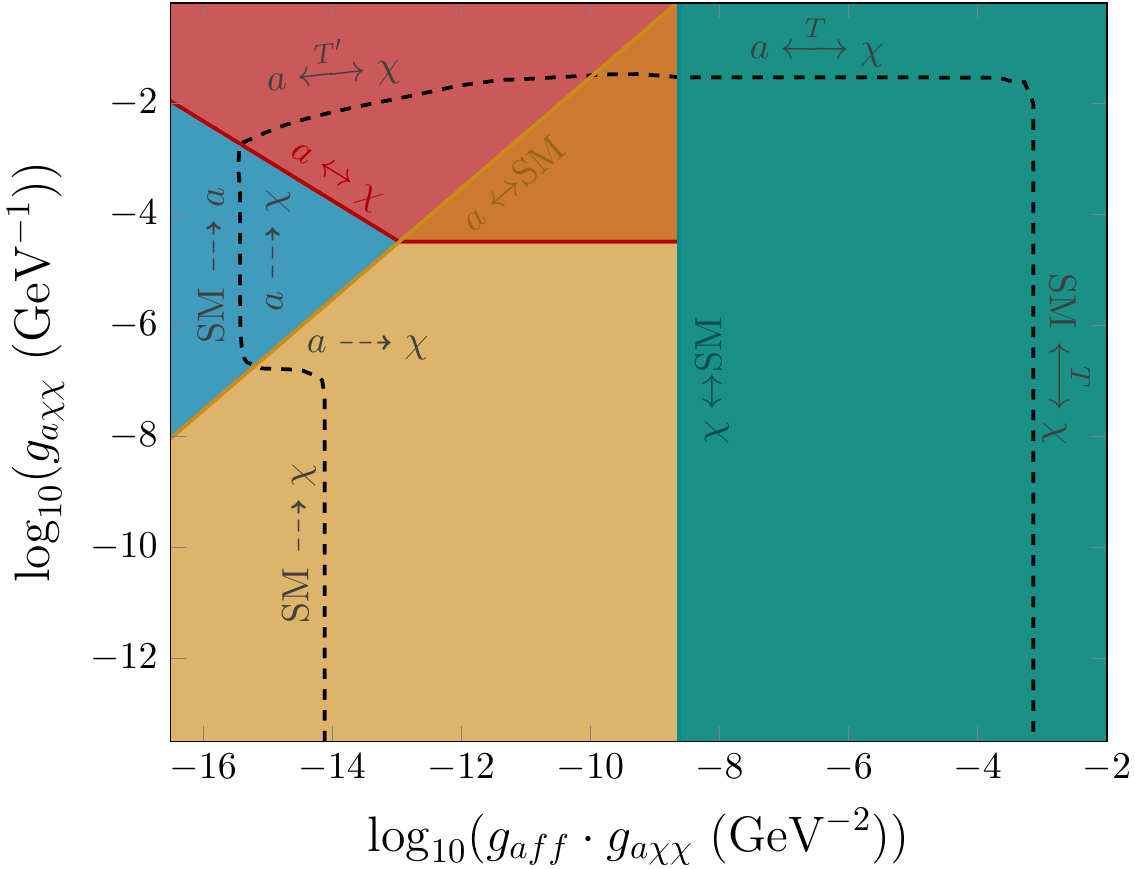}
	\caption[]{ALP-DM coupling $g_{a\chi\chi}$ as a function of the product of the ALP-DM and the ALP-fermion coupling, $g_{a\chi\chi}\cdot g_{aff}$, for $m_\chi=10$~GeV and $m_a=1$~GeV. The dashed line corresponds to the combination of couplings which give the correct DM relic density $\Omega_{\mathrm{DM}}h^2=0.12$ as measured by the Planck telescope.}\label{fig:phase}
\end{center}
\end{figure}
Solving this set of equations for a range of hidden sector and connector couplings gives the phase diagram in figure~\ref{fig:phase} with the dashed line indicating the combination of couplings which lead to the observed DM relic density for $m_\chi=10$~GeV and $m_a=1$~GeV. For a particular theory with some particle content and various interactions it is generic that different types of mechanisms can set the DM relic abundance depending on which type of interaction is more important; our model thus gives rise to a rich phenomenology. In fact, as pioneered in \cite{4ways} and later shown in \cite{Hambye:2019dwd}, the massive mediator case leads to five different dynamical DM production mechanisms. In this work we will focus on scenarios where the DM particles are thermally decoupled from the SM bath. At the lower left of the phase diagram in figure \ref{fig:phase}, corresponding to small $g_{a\chi\chi}$ and a small product $g_{aff}\cdot g_{a\chi\chi}$, the DM particles are not in equilibrium, neither with the SM particles nor with the ALPs. However, the ALP fermion coupling is large enough for the ALPs and the SM to equilibrate, $a \stackrel{T}{\longleftrightarrow}$ SM. DM is produced directly from collisions of SM fermions, $f\bar f \to \chi \bar \chi$. This regime is therefore called \emph{freeze-in from SM}. Increasing $g_{a\chi\chi}$ while keeping the product $g_{aff}\cdot g_{a\chi\chi}$ fixed, $aa \dashrightarrow \chi \bar \chi$ becomes the dominant process. Being conceptually very similar to the previous mechanism, this mechanism is dubbed \emph{freeze-in from the mediator}. If we continue decreasing the ALP fermion coupling $g_{aff}$, ALP-SM interactions become too rare for both sectors to thermalize. Yet, the amount of produced ALPs is sufficient to obtain the correct DM relic density via the chain of sequential out-of-equilibrium productions SM$\to$ ALPs and $aa \to \chi \bar \chi$; DM is produced via \emph{sequential freeze-in}. However, as the ALPs are not in kinetic equilibrium, the unintegrated Boltzmann equations describing the evolution of the ALP's phase space distribution should be studied for a reliable quantitative analysis \cite{Belanger:2020npe}, rendering the calculation of the final DM abundance non-trivial. By further increasing the hidden sector coupling $g_{a\chi\chi}$, at some point, the DM-ALP interaction becomes so strong that they will thermalize among each other. In particular, the DM and the ALPs will share a common temperature $T^\prime$ which is different from and in general much smaller than the photon temperature $T$. Hence, we need to evaluate their interaction cross-section at their common temperature. More precisely, we calculate the hidden sector temperature $T^\prime$ from its energy density $\rho^\prime$ which can be obtained by solving an additional Boltzmann equation describing the amount of energy transfer from the SM bath to the hidden sector. Hence, we numerically need to solve a system of three coupled (stiff) differential equations. The DM relic density is set by $aa\leftrightarrow \chi \bar \chi$ interactions and the mechanism resembles ordinary freeze-out but occurs at a different temperature. We therefore call this regime \emph{decoupled freeze-out (DFO)}. Finally, on the right side of the phase diagram all three sectors share a common temperature and DM is produced via thermal freeze-out.
\section{Phenomenological Implications}
\begin{figure}[t]
\begin{center}
	\includegraphics[scale=0.3]{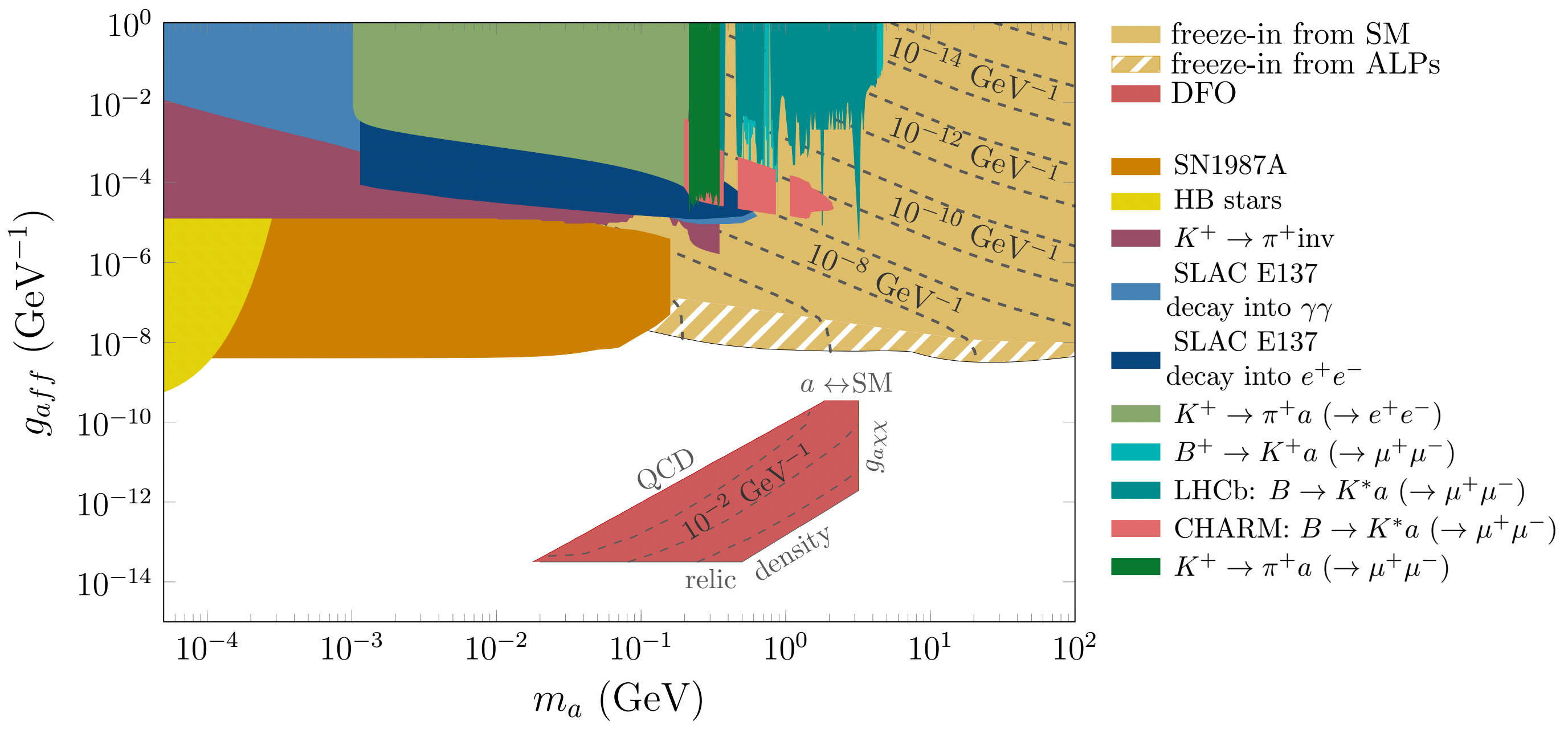}
	\caption[]{ALP-fermion coupling $g_{aff}$ as a function of $m_a$ for lines of constant hidden sector coupling, $g_{a\chi\chi}$ (plotted in dashed grey), which give the observed DM relic density via freeze-in from the SM (yellow region), freeze-in from ALPs (dashed yellow region) and decoupled freeze-out (red region). We fixed the ratio $m_\chi/m_a=10$. We have included the relevant constraints on our ALP model on the connector coupling $g_{aff}$ in this parameter region.}\label{fig:constraints+fi+fodds}
\end{center}
\end{figure}
Last not least, we explore the phenomenological implications for our model by comparing it to experiments. In figure~\ref{fig:constraints+fi+fodds} we show the ALP fermion coupling as a function of the ALP mass with the lines of constant hidden sector couplings $g_{a\chi\chi}$ which give the correct relic density in the freeze-in and DFO regimes. We keep the ratio between the mass of the ALP and the mass of the DM $m_\chi/m_a=10$ fixed. We also show the exclusions from beam dump, astrophysics and flavour physics which we calculated for our specific model, namely where the ALP only couples to SM fermions at tree-level and ALP-gauge boson couplings are induced by fermion loops. Astrophysical constraints coming from the neutrino burst from SN1987A or the ratio of the number of horizontal branch stars to red giants in globular clusters rely on energy loss arguments \cite{Chang:2018rso,kevmass}. The SLAC electron beam dump experiment searched for axions and observed no events \cite{SLAC}. For the flavour constraints, we derive bounds for the most constraining channels: $K^+\to\pi^+a (\to {\mathrm{inv.}})$ from NA62, $K^+\to\pi^+a(\to\mu^+\mu^-)$ and $K^+\to\pi^+ a( \to e^+e^-)$ from NA48/2, $B^+\to K^+a(\to\mu^+\mu^-)$ from LHCb and $B \to K^*a(\to\mu^+\mu^-)$ from LHCb and CHARM.
The freeze-in mechanism is possible in a large parameter space in the $g_{aff}-m_a$ plane. The connector coupling is comparably large since the ALPs and the SM particles are in equilibrium and is therefore likely in reach for detection experiments. For $g_{aff}$ above $\sim 10^{-6}-10^{-5}$~GeV$^{-1}$ and $m_a$ below a few GeV, our model is ruled out by flavour constraints and the electron beam dump experiment at SLAC. For masses below $\sim 0.2$~GeV, the remaining parameter space is covered by the SN constraint. In the DFO regime, contrarily to the freeze-in regimes, $g_{a\chi\chi}$ is large to ensure equilibrium among the hidden sector particles. However, it appears to be extremely challenging to test the DFO region with collider searches for ALPs. Yet, due to the tiny coupling between the SM particles and the ALP, the ALP is long-lived. Since the ALPs are abundantly produced along with the DM particles their decay has important implications for the cosmological history at the time of big bang nucleosynthesis, resulting in big bang nucleosynthesis constraints.
\section{Conclusion}
Our model of an ALP acting as a mediator between the SM particles and the DM particles engenders a rich phenomenology: Depending on the ALP and DM masses and couplings, various mechanisms which reproduce the DM relic density measured today are possible. The calculation of the relic density in the DFO region is particularly non-trivial. We performed a detailed calculation of this region in this ALP mediated DM model.  Note that here we considered IR-dominated contributions to the final relic density. In general, depending on the reheating temperature, there could also be UV-dominated contributions from higher dimensional operators since the Yukawa-type terms merely represent effective interactions, see \cite{ALPmediator}. However, these do not change the qualitative conclusions. To apply existing constraints to our model we have revisited astrophysical and collider constraints on our ALP. We look forward to seeing the impact of future experimental results on our model.
\section*{Acknowledgments}
I am grateful to the organisers of the conference, to my supervisor Aoife Bharucha and my collaborators Felix Br\"ummer and Nishita Desai. This work was supported by the doctoral school ED352 and has received funding from the Excellence Initiative of Aix-Marseille University - A*MIDEX, a French ``Investissement d'Avenir'' programme (AMX-19-IET-008-iPhU).

\end{document}